Peculiar black hole accretion rates in AGN with highest star formation rates in the universe

David Garofalo[1] & Ektoras Pouliasis[2]

1. Department of Physics, Kennesaw State University
2. IAASARS, National Observatory of Athens, Ioannou Metaxa and Vasileos Pavlou GR-15236, Athens, Greece

Abstract

Pouliasis et al (2022b) explored star formation rates, black hole accretion rates, and stellar mass of active galaxies at redshift above 3.5, uncovering a leveling off of the star formation rate at high stellar mass, which they consider to be evidence of AGN feedback. Their data shows that as AGN approach the flattening of the curve in the star formation rate – stellar mass plane, the accretion rates begin to drop. We describe the nature of the AGN feedback responsible for this in terms of powerful FRII jets enhancing star formation rates but eventually also triggering a shift in accretion from near-Eddington rates to advection dominated. These systems are on the cusp of a dramatic transition where the active galaxy goes from strong enhancement to large suppression of star formation in a way that produces the steeper slope for radio AGN at low redshift compared to radio AGN at higher redshift and to jetless AGN. We argue, therefore, that the data of Pouliasis et al constitute the high redshift objects predicted by Singh et al (2021) that connect to the low redshift behavior of radio AGN shown in Comerford et al (2020).

1. Introduction

From surveys such as MaNGA (Blanton et al 2017), GOODS (Giavalisco et al 2004), COSMOS (Scoville et al 2007) , and now with JWST (Gardner et al 2006), it has become possible to deepen our understanding of the effect that black holes have on star formation over cosmic time. We are arguably approaching a time when it will be possible to piece together the nature of the co-evolution of black holes with galaxies starting from hundreds of millions of years after the Big Bang to the present time, shedding light on the details of their feedback and the origin of the black hole scaling relations (Magorrian et al 1998; Ferrarese & Merritt 2000; Gebhardt et al 2000).

Comerford et al (2020) found that jetted AGN at low redshift distribute themselves further rightward and with greater slope than non-jetted AGN on the star formation rate – stellar mass plane (SFR-SM plane). Singh et al (2021) showed that such difference can be understood as the late stage evolution of jetted AGN in more dense environments where powerful jetted AGN feedback operates. In so doing, Singh et al (2021) predicted the behavior of accreting black holes at higher redshift. Pouliasis et al (2022b) added 89 active galaxies (AGN) above redshift 3.5 to the star formation rate – stellar mass plane and argued for AGN feedback at high stellar mass possibly suppressing star formation. We analyze this work in the context of the AGN feedback scenario described in Singh et al (2021), and show how to connect them to the lower redshift radio loud

AGN analyzed by Comerford et al (2020). Our focus is on the decreasing specific black hole accretion rates in Pouliasis et al (2022b) at high stellar mass.

In Section 2 we describe the data in Pouliasis et al (2022b), the ideas in Singh et al (2021) that allow us to interpret this data and the connection to the low redshift data in Comerford et al (2020). In Section 3 we conclude.

2. Discussion

2.1 Data

Pouliasis et al. (2022b) compiled a large number (149) of high redshift AGN selected in the Chandra COSMOS Legacy survey, eROSITA Final Equatorial Depth Survey and XMM-XXL North (Marchesi et al 2016; Salvato et al 2021; Pouliasis et al 2022a). They used the X-CIGALE (Yang et al 2020; Yang et al 2022) SED fitting algorithm, that has been used widely in the literature (e.g. Pouliasis et al 2020; Koutoulidis et al 2022), to derive the AGN host galaxy parameters. Applying several quality criteria, they ended up with 89 sources with reliable star-formation rate and stellar mass measurements. Pouliasis et al (2022b) have found what appear to be the critical transition objects predicted by Singh et al (2021), which amounts to AGN at high redshift ($z > 3.5$) with very high SFR and large SM whose specific black hole accretion rates are lower.

As we will describe in Section 2.2, the AGN whose feedback generates the highest SFR also modify their accretion away from near Eddington values and towards advection dominated ones (Singh et al 2021). The data from Pouliasis et al (2022b) that we plot in Figure 1 displays this trend. While all the objects in Figure 1 accrete above the theoretical threshold that divides radiatively efficient disks from advection dominated ones at log $(dM/dt)_{Edd}$ = -2 (Shakura & Sunyaev 1973; Narayan and Yi 1995), where $(dM/dt)_{Edd} = (dM/dt)/(dM/dt)_{max}$ and $(dM/dt)_{max}$ is the Eddington-limited accretion rate, the blue objects are characterized by $(dM/dt)_{Edd} > 0.5$ while the green ones by $(dM/dt)_{Edd} < 0.5$. We find that a decrease in specific black hole accretion rates begins to set in at higher SFR and higher SM. This is the prediction of Singh et al (2021) that we describe in the next section.

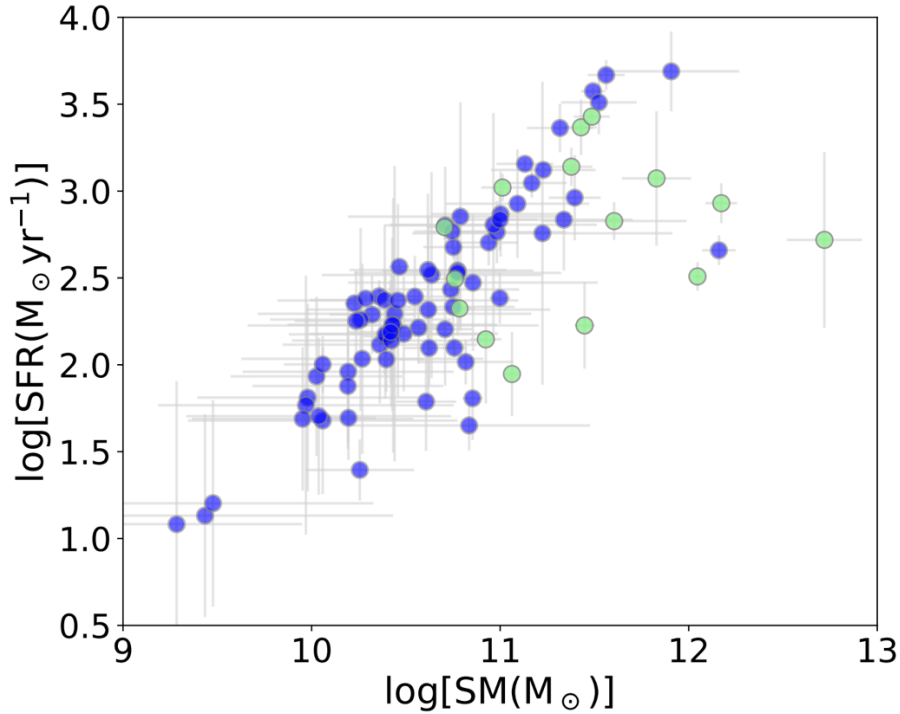

Figure 1: The AGN of Pouliasis et (2022b). The blue are characterized by $(dM/dt)_{Edd} > 0.5$ while the green by $(dM/dt)_{Edd} < 0.5$.

2.2 Theory

While the majority of black holes of any mass are unlikely to end up in a counterrotating accretion configuration, the key element in the formation of powerful Jetted AGN is a merger that funnels cold gas into its nucleus that settles in counterrotation around a spinning black hole (Garofalo, Evans & Sambruna 2010). In order for such configurations to be stable, the total angular momentum of the disk must not be large relative to that of the black hole (King et al 2005). While very massive black holes can be formed in isolated environments, they are less likely to end up in a counterrotating accretion configuration. The details of this can be found in Garofalo, Christian & Jones, 2019. The take away point is that counterrotating black holes, if formed, tend to be less massive in isolated environments, on average, and more so on average in cluster environments. And since jet power is larger for more massive black holes, the jet powers in cluster environments are on average larger. This idea lies at the heart of the formation of jetted AGN in the paradigm that is described in Figure 2. The conditions for the formation of counterrotating accretion around black holes are stringent which makes jetted AGN the minority (i.e. it explains the radio loud/radio quiet dichotomy). Hence, most mergers lead to corotation. Corotation is also dominant in AGN that are triggered by secular processes. These characteristics allow us to understand the evolution of AGN in all environments as shown in Figure 2, which shows different phases in the time evolution of AGN. Our goal is to describe only in a detailed enough way to understand the interpretation of the data in Pouliasis et al (2022b).

In the upper left panel we see a black hole that is spinning at 90% of its maximum rate and is accreting via a thin disk in corotation around the black hole. This kind of AGN can result via a merger (e.g. a radio quiet quasar) or by secular processes (e.g. a narrow line Seyfert 1). Rapidly spinning corotating black holes suffer jet suppression and therefore do not produce jets (Garofalo & Singh 2016; Ponti et al 2012; Garofalo, Evans & Sambruna 2010; Neilsen & Lee 2009). Such AGN simply spin their black holes up to the maximum value until the accretion fuel is exhausted. Because the innermost stable circular orbit is closer to the black hole for higher black hole spin, there is greater energy reprocessed via the thin disk which leads to a stronger disk wind shown in red. This disk wind has a relatively mild SFR suppression ability compared to jets. The triggers of such AGN also lead to the onset of star formation which is shown in the left column.

The top right panel of Figure 2, instead, shows the evolution of the average black hole triggered into counterrotation by a merger in an isolated environment. The combination of a rapidly spinning black hole and counterrotation maximizes both the Blandford-Znajek jet (BZ) as well as the Blandford-Payne jet (BP). This leads to an FRII jet. We refer to the combination of FRII jet morphology and a radiatively efficient disk as an FRII HERG for high excitation radio galaxy. The time evolution is characterized by physics that is not ad-hoc or based on additional assumptions but which involves the consequences of accretion, which are a black hole that spins down and then up again in corotation. FRII jets in such environments tend to produce weaker feedback compared to richer environments because the black holes are least massive in field environments (as discussed above, the most massive black holes in isolated environments are less likely to end up in counterrotation). The FRII jet enhances the star formation rate but otherwise the host galaxy and AGN evolve in a way that is not strongly affected by black hole feedback.

In the lower panels of Figure 2, by contrast, we see the effect of black holes triggered by mergers in increasingly denser environments where the likelihood of a more massive black hole accreting in counterrotation increases. In the lower panel of Figure 2 we begin in the same way as in the field environment but the difference is that black holes can be more massive in groups. As a result, the FRII jet has an effect on the accreting material that manifests itself in a change at late times from a radiatively efficient thin disk to an advection dominate accretion flow (ADAF).

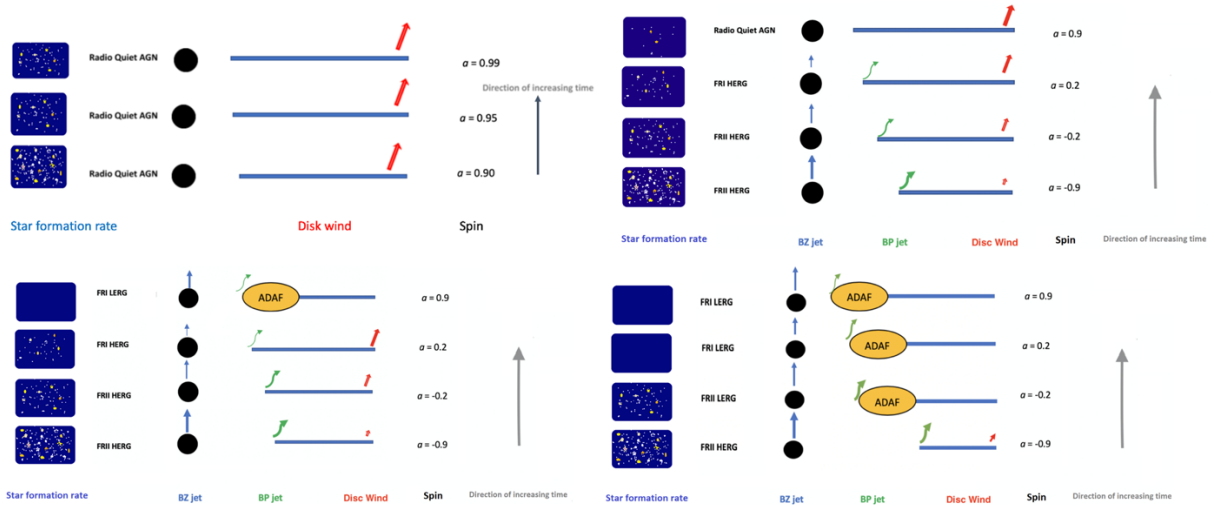

Figure 2: The average evolution of accreting black holes triggered via mergers and/or secular processes (upper left panel), via mergers into counterrotating configuration in isolated environments (upper right panel), via mergers into counterrotating configurations in intermediate density environments or groups (lower left panel), via mergers into counterrotating configurations in rich cluster environments (lower right panel).

The transition through zero spin can change the direction of the jet (Garofalo, Joshi et al 2020) and this will have a strong negative impact on the SFR (The Roy Conjecture - see Garofalo & Mountrichas 2022). Hence, the SFR begins to drop appreciably at late times when the black hole is already rapidly spinning in corotation with its accretion disk.

In the densest environments, the extreme values that are possible on average for black hole mass makes the counterrotating configuration particularly powerful in its feedback. As a result, the disk rapidly evolves away from a thin configuration and into an ADAF, which means the accretion rate has crossed the threshold $10^{-2}$ that of Eddington. The timescale for this change to begin is only about 4 million years. Hence, within just a few million years after the triggering of the massive black hole in counterrotation, the accretion disk is moving away from a radiatively efficient thin disk and into an ADAF which means the rate of accretion transitions from the Eddington rate to 0.01 that value within 4 million years (Antonuccio-Delogu & Silk 2010; Garofalo, Evans & Sambruna 2010). Such systems go from the label FRII HERG to FRII LERG where the latter refers to low excitation radio galaxy. The FRII jet remains (although the conditions for the BZ and BP jets are weaker) because it takes $8 \times 10^6$ years for a rapidly spinning black hole to spin down to zero via a thin disk (e.g. Rayne & Thomas 2005; Kim et al 2016) and 4 million years for the transition into an ADAF means the object is still characterized by counterrotation. While the accretion rate is dropping, the star formation continues to be enhanced because there is a direct connection between FRII jets and SFR enhancement (e.g. Kalfountzou et al 2014; Garofalo et al 2016). FRI jets will form once the spin crosses the 0.2 spin threshold (Garofalo, Evans & Sambruna 2010). As in groups, here too the FRI jet is likely to be tilted with respect to the previous FRII jet and this has the same qualitative feedback effect on star formation, except more dominant because the black hole mass is greatest in such environments. It is here that SFR are strongly suppressed and this is captured in the left column of the right hand side of the lower panel of

Figure 2. Of crucial importance for this work is the transition during the FRII jet phase from a HERG to a LERG.

Singh et al (2021) have plotted the average behavior of counterrotating systems in the SFR-SM plane, showing the change in paths through zero black hole spin. For simplicity, they start the AGN on the radio quiet line and describe the evolution in the different environments. We reproduce their result in Figure 3. Briefly, the three green to pink paths represent the evolution in the three different environments labeled in yellow. The green path is the counterrotating phase while the pink path is the corotating phase. The feedback goes from positive via FRII jets to negative via FRI jets. The data of Pouliasis et al (2022b) makes the path of clusters, and the transition point labeled B, of importance for this work. The green path is the FRII, SFR enhancement phase that ends at the pink path, indicating a transition into the SFR suppression phase due to the tilted FRI jet. Given the data shown in Figure 1, we predict that the green objects of Figure 1 are at the end stage of paths like the one shown in Figure 3 but that have starting points that do not begin on the radio quiet AGN line.

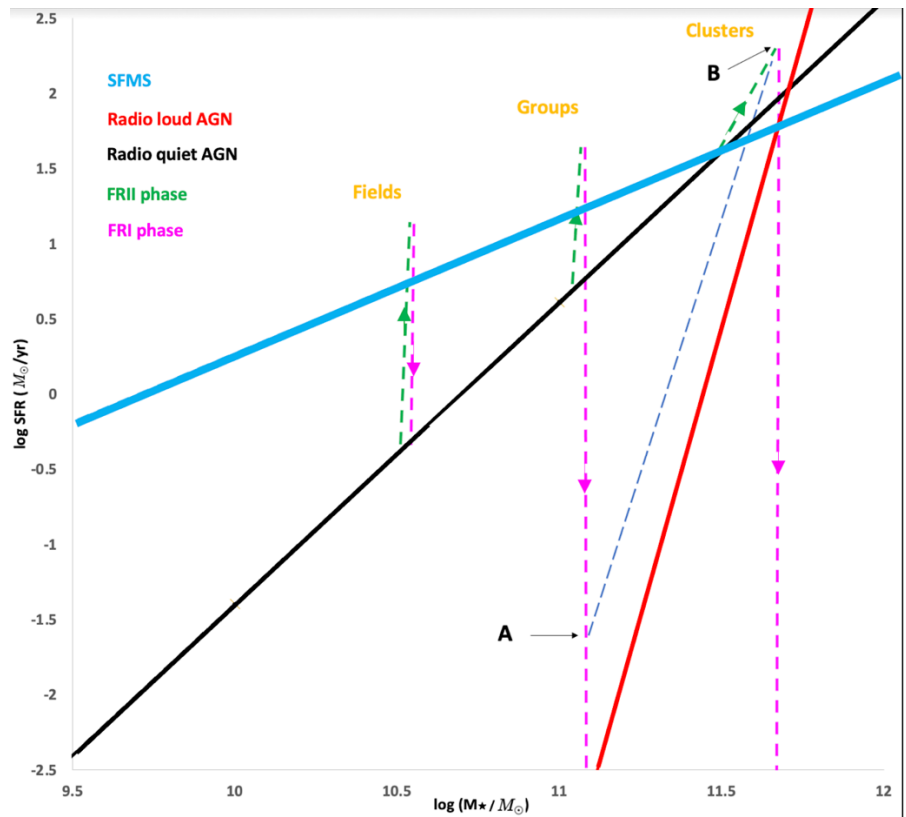

Figure 3: The paths of counterrotating accreting AGN described in Singh et al (2021). The figure is a modification on the theme from Garofalo & Mountrichas (2022). Points A and B explain the slope of the radio loud AGN.

In Figure 4 we place the objects of this study in context by adding the predicted radio quiet AGN of Mountrichas et al (2021), the radio loud AGN at low redshift of Comerford et al (2020), and the average for the extreme radio galaxies of Ichikawa et al (2021). From the perspective of the model described in Figure 2 and applied to the SFR-SM plane in Figure 3, we understand that the green objects in Figure 4 will become FRI LERGs. When this happens, they will provide an SFR suppression at low accretion rates, which means they increase their SM values slowly. As a result, they move mostly downward to become like the red objects at low redshift. Although some of these green objects are transitioning through zero black hole spin as described by the evolution in group environments, we should also expect to see some FRII jets among this green population as captured by the lower right panel of Figure 2, although they may be of low power due to low black hole spin. Additionally, we predict that the jetted AGN in red in Figure 4 are dominated by ADAF accretion and thus are LERGs. In fact, they are predicted to be FRI LERGs. As described in Garofalo & Mountrichas (2022), the purple objects should be dominated by jetless AGN and the reason that purple struggles to dominate at low SFR is because such systems fail to produce the powerful FRII jets that are responsible for the late LERG states predicted for the green objects. The yellow point describes the average of jetted AGN with high accretion rates, which constitutes evidence that we indeed have powerful jets among this class of high accreting, high star formation, but generally low stellar mass systems.

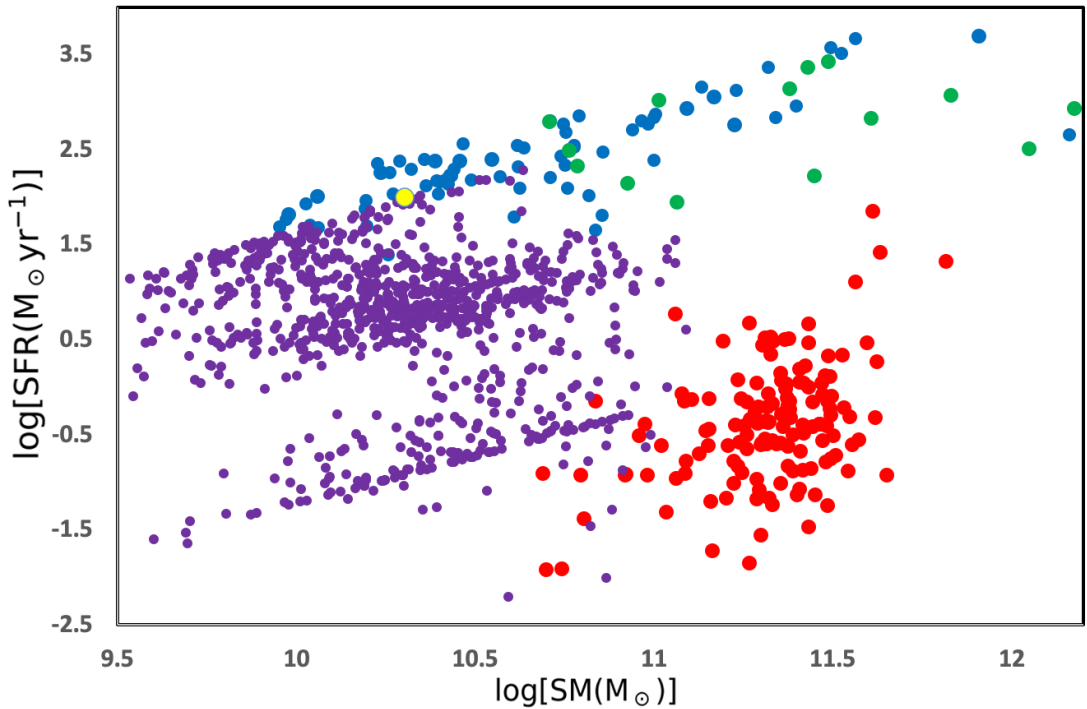

Figure 4: The high SFR objects of Pouliasis with the higher accreting subgroup in blue and the lower accreting ones in green. The lower redshift jetted AGN of Comerford et al (2020) are in red, the AGN of Mountrichas et al (2021) that were predicted to be mostly jetless by Garofalo & Mountrichas (2022) are in purple, and the average for the high accretion rate and jetted objects of Ichikawa et al (2021) is in yellow.

The difference in accretion among the objects of Pouliasis et al (2022b) is relevant for the model, which argues for the absence of a strong connection between SFR and accretion rate in AGN with powerful jets due to the strong jet feedback effect in these systems. FRII HERGs as well as FRII LERGs produce a positive feedback effect on the SFR. FRI HERGs (which we do not focus on here) and more importantly FRI LERGs tend to have the opposite effect on the SFR. Hence, the prediction is that SFR will not show any correlation with accretion rate or excitation level. In order to explore this, we plot SFR versus Eddington specific black hole accretion rates for the objects of Pouliasis et al (2022b) in Figure 5. We do not see any correlation. Because FRII HERG and FRII LERG tend to form mostly in the richest environments, we predict that Figure 5 should be environment dependent. As the environment density drops, we predict that the number of green objects would decrease more than the number of blue objects. This is because in field environments one expects fewer FRII HERG and these tend not to evolve into LERG.

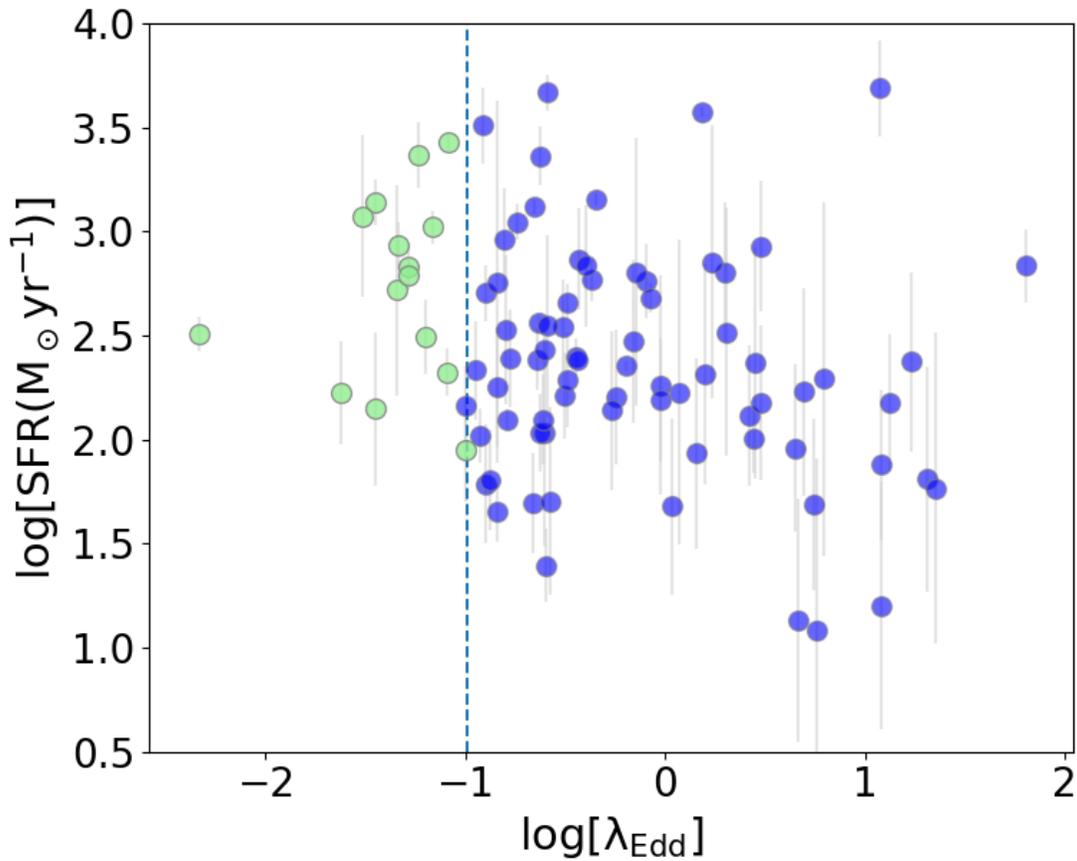

Figure 5: The AGN of Pouliasis et al (2022b) as a function of the Eddington accretion rate showing absence of correlation.

3. Conclusions

The drop in accretion rates as one approaches the levelling off in SFR at high stellar mass and high star formation rates uncovered by Pouliasis et al (2022b) appears to be a subtle but important constraint on the nature of AGN feedback. At the highest SFR, the feedback from AGN that sets in and that will eventually lower the SFR is still in the future for the green objects of Figure 1 and Figure 4. The telltale signature of this from the model perspective is the lower accretion rates despite high SFR values. We have provided context for these objects by including previously explored subclasses of AGN on the SFR-SM plane and are able to make predictions. In closing, we caution that SFR-SM is not the most direct way to explore the implications of the model. This is because the model as seen in Figure 2 is more fundamentally about SFR versus cluster richness (CR). We therefore propose that SFR-CR may be the ideal parameter space for understanding AGN feedback. The model would more precisely describe the distribution of SFR as a function of CR and why low SFR, in particular, do not occur in low CR environments to the degree they do in high CR environments. This is, of course, tied to mergers and the subpopulation of mergers that experience counterrotating accretion flows of cold gas into their nuclei. There is an expectation that LINER AGN and FRI LERG are the end stages of AGN that were not triggered with counterrotating disks and that were, respectively. In other words, in the SFR-CR plane, we can understand the dominant contribution of high excitation systems (whether jetted or not) at low CR values, with an increase in the low excitation systems as CR increases, specifically with jetted AGN (i.e. LERG). High SFR and low CR is where we predict to find the objects that Foschini has been discussing over the last decade (i.e. gamma narrow line Seyfert 1 and narrow line Seyfert 1 – Foschini et al 2015). Gamma narrow line Seyfert 1 would be jetted AGN that form from black holes whose spins are intermediate and whose accretion disks settle into corotation. The connection between jetted and non-jetted narrow line Seyfert 1 AGN is where the model sheds light on the jet suppression phenomenon in spirals. A key limitation of this model, in this context, is its weakness in determining the spin value associated with jet suppression in corotating accretion states (see Garofalo & Singh 2022), which is a key element of accreting black holes across the mass scale (e.g. X-ray binaries).

5. Acknowledgments

The research leading to these results has received partial funding (EP) from the European Union's Horizon 2020 Programme under the AHEAD2020 project (grant agreement n. 871158).